\documentclass[preprint]{rmaa}

\usepackage[dvips]{color}

\title{Hypercritical accretion onto a magnetized neutron star surface: a numerical approach}

\author{
  C. G. Bernal,\altaffilmark{1} 
  W. H. Lee,\altaffilmark{1}
  and Dany Page\altaffilmark{1}}

\altaffiltext{1}{Instituto de Astronom\'{\i}a, Universidad Nacional Aut\'onoma de M\'exico.}

\shortauthor{Bernal, Lee, \& Page}
\shorttitle{Hypercritical Accretion onto a magnetized NS Surface}

\fulladdresses{
\item C. G. Bernal, W. H. Lee, D. Page: Instituto de Astronom\'\i{}a, Universidad Nacional Aut\'onoma de M\'exico,
  Apdo. Postal 70 - 264, Ciudad Universitaria, M\'exico D.F, CP 04510
  (bernalcg,wlee,page@astroscu.unam.mx).}

\listofauthors{C. G. Bernal, W. H. Lee \& D. Page}
\indexauthor{Bernal, C. G.}
\indexauthor{Lee, W. H.}
\indexauthor{Page, D.}

\abstract{
The properties of a new--born neutron star, produced in a core--collapse supernova, can be strongly affected by the possible late fallback
which occurs several hours after the explosion.
This accretion occurs in the regime dominated by neutrino cooling, explored initially in this context by  \citet{Chevalier:1989jk}.
Here we revisit this approach in a 1D spherically symmetric model and carry out numerical simulations in 2D in an accretion column onto a neutron star considering detailed microphysics, neutrino cooling and the presence of magnetic fields in ideal MHD. We compare our numerical results to the analytic solutions and explore how the purely hydrodynamical as well as the MHD solutions differ from them, and begin to explore how this may affect the appearance of the remnant as a typical radio pulsar.}

\resumen{La acreci\'{o}n sobre una proto-estrella de neutrones en las horas que siguen al colapso del n\'{u}cleo de una estrella masiva que le dio origen puede afectar sus propiedades observables. Este fen\'{o}meno se da en el regimen denominado hipercr\'{\i}tico (Chevalier 1989), donde el enfriamiento por neutrinos es crucial para la evoluci\'{o}n termodin\'{a}mica. En este trabajo presentamos un estudio en este contexto en una dimensi\'{o}n con simetr\'{\i}a esf\'{e}rica y llevamos a cabo simulaciones num\'{e}ricas en dos dimensiones dentro de una columna de acreci\'{o}n sobre una estrella de neutrones. Consideramos procesos microf\'{\i}sicos detallados, enfriamiento por neutrinos y la presencia de campos magn\'{e}ticos en la aproximaci\'{o}n de magnetohidrodin\'{a}mica ideal. Comparamos nuestros resultados num\'{e}ricos con las soluciones anal\'{\i}ticas e investigamos como las soluciones, tanto hidrodin\'{a}micas como magnetohidrodin\'{a}micas, difieren de \'{e}stas. Iniciamos tambi\'{e}n una exploraci\'{o}n de como este proceso puede afectar la aparici\'{o}n del remanente como un pulsar t\'{\i}pico en el radio.}

\addkeyword{Accretion}
\addkeyword{Hydrodynamics}
\addkeyword{Magnetic Fields}
\addkeyword{Supernovae: individual (SN1987A)}
\addkeyword{Stars: Neutron}

\begin{document}
\maketitle

\section{Introduction}
\label{sec:intro}

The neutrino signal \citep{Hirata:1987jk,Bionta:1987mz} detected from the supernova SN1987A 
clearly demonstrated the birth of a neutron star \citep{Burrows:1986gf}.
Identification of the progenitor as the blue supergiant Sanduleak $-69^\circ 202a$ \citep{Gilmozzi:1987xy}
and modeling of the early light curve \citep{Hillebrandt:1987qv,Shigeyama:1987nr}
proved that the supernova resulted from the core--collapse of a massive, $\sim 20$~M$_\odot$, star.
However, to date, there is no evidence for the presence of a pulsar, or even a quiet neutron star, in the remnant
(see, e.g., discussion in \citealt{Haberl:2006bh} and \citealt{Shternin:2008lq}).
Several solutions to this dilemma have been proposed as, e.g., the delayed collapse of the neutron star into a black-hole
\citep{Ellis:1996dq,Brown:1994rr}
or a delayed turn-on of the pulsar \citep{Michel:1994cr,Muslimov:1995sf}.
The latter case is just an extreme case of the more mundane possibility that the neutron star is weakly magnetized and/or slowly rotating, 
resulting in a spin-down energy that is low enough so as to be undetectable.

In recent years, several  measurements of pulsar masses point toward large values approaching, or even exceeding, 2 M$_\odot$ 
(see, e.g., \citealt{Freire:2008wd}) which would strongly disfavor the black-hole explanation.
On the other side, timing of radio quiet compact stars in young supernova remnants,
usually dubbed  CCOs ("Central Compact Objects",  \citealt{Pavlov:2002fk}) recently unveiled at least three case of weakly 
magnetized young neutron stars \citep{Gotthelf:2008eu}:
PSR J0821-4300 (in the SNR Puppis A) 
with a rotational period $P=112$ ms, an upper limit on its spin-down power
$\dot{E} < 2.3 \times 10^{35}$ erg s$^{-1}$, and a surface dipolar magnetic field strength $B_\mathrm{dip} < 9.8 \times 10^{11}$ G
\citep{Gotthelf:2009oq};
PSR 1E1207.4-5209 (in the SNR PKS 1209-51/52)
with $P=424$ ms, $\dot{E} < 1.3 \times 10^{32}$ erg s$^{-1}$, and $B_\mathrm{dip} < 3.3 \times 10^{11}$ G \citep{Gotthelf:2007qe},
and finally 
PSR J1852+0040 (in the SNR Kes 79) 
with $P=424$ ms, and measurements of $\dot{E} = 3.0 \times 10^{32}$ erg s$^{-1}$, and $B_\mathrm{dip} = 3.1 \times 10^{10}$ G \citep{Halpern:2010nx}.
The last two of these hence have an energy output well below the 0.2-10.0 keV luminosity of the SN 1987A remnant, 
$L<5.7\times 10^{34}$  erg s$^{-1}$ \citep{Haberl:2006bh}. If the neutron star produced by SN 1987A has similar characteristics
it would presently be undetectable.

\begin{figure}[!t]\centering
  \includegraphics[width=0.75\columnwidth]{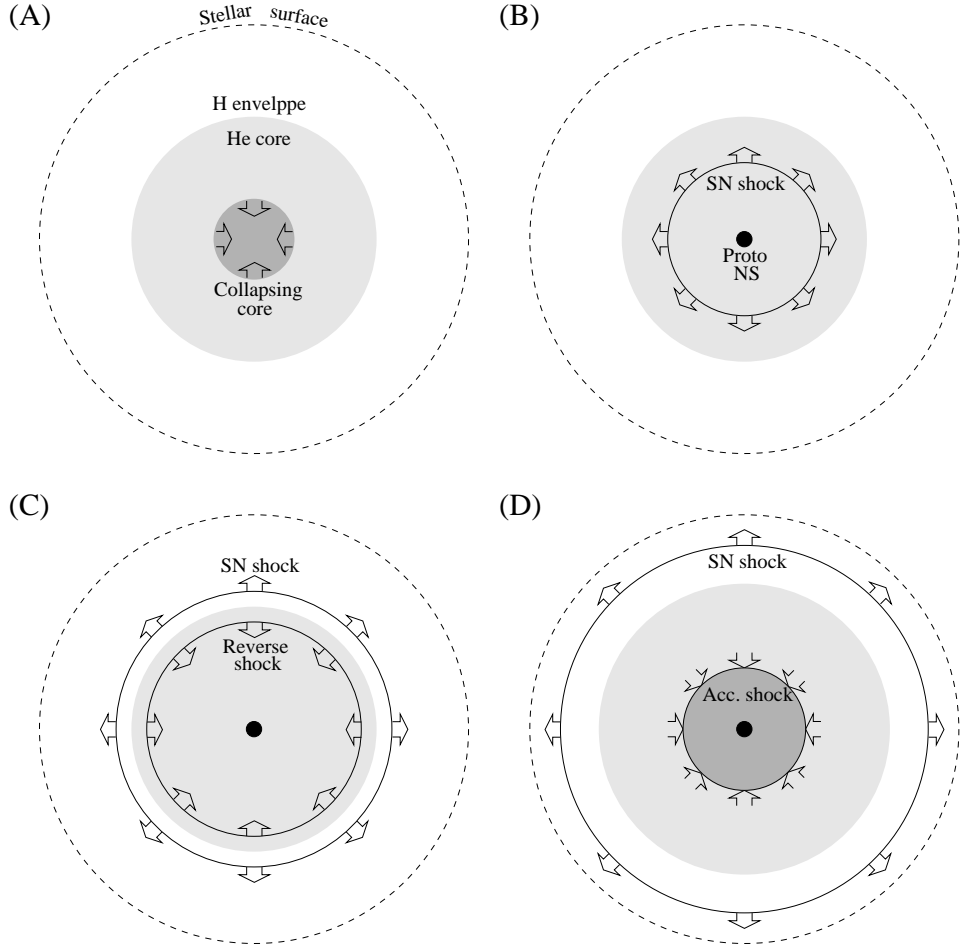}
  \caption{Schematic evolution of a collapsing stellar core. A, the Chandrasekhar--mass Iron core collapses; B, upon formation of a proto--neutron star, the equation of state stiffens and produces a bounce, launching an outwardly propagating shock; C, a reverse shock is formed at the He--H interface in the stellar envelope, moving back towards the neutron star; D, an accretion shock is established, interior to which an atmosphere close to hydrostatic equilibrium in the neutrino--cooled regime deposits mass and energy onto the neutron star in the hours following the explosion.}
  \label{Fig:scenario}
\end{figure}

In the present paper we consider the scenario in which the initial magnetic field of the new-born neutron star is strongly modified by
a phase of late, and intense, accretion, occurring a few hours after the initial explosion \citep{Geppert:1999rz}.
When a massive star explodes as a supernova, following the core--collapse scenario 
\citep{Woosley:2005ty,Mezzacappa:2005jk,Janka:2007xy}, a large fraction of its mass expands freely and interacts with the interstellar medium.
However, the central compact remnant also interacts with the inner envelope through its gravitational field. 
In Type II SNe (see Figure~\ref{Fig:scenario}), the initial core--collapse (panel A) produces a proto--neutron star when the equation of state stiffens close to nuclear density. A high--velocity ($10^{4}\,\mathrm{km\,s^{-1}}$) expansive shock then starts moving outward (panel B).
Flow lines bifurcate and some part of the matter falls back onto the central remnant. The rest is unbound and ends up being ejected in the explosion. 
This scenario produces a low density region in near free fall between the surface of the compact object and the extended atmosphere in near hydrostatic equilibrium that has formed. 
In case the progenitor star had a low density envelope surrounding the He core a reverse shock (panel C) decelerates the matter
and causes a late fallback onto the compact object, depositing great amounts of matter onto the surface of the
new--born neutron star in the hours following the explosion (panel D).
Following the ideas of 
\citet{Blondin:1986sf}, \citet{Chevalier:1989jk}, \citet{Houck:1991rz}, and \citet{Brown:1994rm}
about the accretion of
matter onto compact objects, it is possible to develop an analytical model of
accretion following core--collapse, and particularly in the case of SN1987A. One of the salient features of this analysis is that the gas, being quite dense, is unable to cool by photon emission, and the mass accretion rates are highly super--Eddington in that sense. However, at sufficiently high temperatures, cooling through neutrinos sets in, mostly through pair annihilation and pair capture onto free nucleons, and given their much lower interaction cross section with matter, they are able to remove enough energy from the flow for accretion to take place. This regime is usually termed \textquotedblleft hypercritical\textquotedblright\ accretion, and is common in the inner collapsing stellar cores and is likely to drive the central engines of Gamma Ray Bursts \citep{Lee:2007zl}.
 With this model it is formally possible to obtain the radial position of the shock as a function of the mass accretion rate from fallback, assuming steady state in spherical symmetry, as well as the structure of the envelope. 

\citet{Chevalier:1989jk} and \citet{Houck:1991rz}
computed such solutions in the context of SN1987A. Here we wish to explore the behavior of the flow under more general conditions, and present solutions for an accretion column in two dimensions, which we compare with the analytical scalings. Neutrino cooling is a crucial ingredient in the relevant density and temperature regimes, and we consider it along with a detailed equation of state. In addition and more importantly, we begin to explore the effects of the magnetic field on the accumulation of matter onto the neutron star surface. This is only possible through 2D simulations of the kind shown here, and we make a comparative analysis between the analytical and numerical approaches to consider the submergence of the magnetic field in the crust of the neutron star and the piling up of matter on its surface. Previously, 
\citet{Muslimov:1995sf} and \citet{Geppert:1999rz}
considered how such accretion might delay the switch--on of a pulsar following its formation in one--dimensional calculations, computing the ohmic diffusion time of the magnetic field through the accreted matter. \citet{Fryer:1996rm} studied the two--dimensional accretion dynamics onto new--born neutron stars in the neutrino cooled regime, finding that in some cases, neutrino--driven convection can significantly modify the simple one--dimensional steady state solution.

Here we report on preliminary, two--dimensional numerical calculations which aim to determine if hypercritical, neutrino--cooled accretion  can submerge the magnetic field into the crust of the neutron star, and if it plays an important role in the dynamics in this regime. In \S ~\ref{sec:analytical} we
develop the analytical model of the hypercritical accretion process and calculate
the structure of the envelope for a two dimensional accretion column.
We build a numerical model, based on these analytical consideration, in \S~\ref{sec:numerical}. 
In \S ~\ref{sec:results}, we show numerical results for various configurations including magnetic fields at several accretion rates and present a comparative analysis between the
numerical and analytical solutions for the scenario of SN1987A.
Finally, in \S ~\ref{sec:conclusions} we present some preliminary conclusions.

\section{Analytical Models}
\label{sec:analytical}

As a benchmark against which to compare our numerical simulations, we summarize below the basic results of an
analytical model, based on the one developed by \citet{Chevalier:1989jk}, and adapt them to the case of an accretion column.
The essential assumptions of the model are that the neutron star is at rest within the expanding medium at infinity and
rotation is neglected.
For this analytical approach, we also neglect the effect of a possible magnetic field.
Matter is described by a polytropic equation of state, $P = K  \rho^\gamma$ with an index $\gamma = 4/3$, and is assumed
to evolve adiabatically except at the shock interfaces and close to the neutron star surface where neutrino emission
(through $e^{\pm}$ pair annihilation) assures that the accretion energy is properly removed from the system.

\subsection{The Initial Late--Accretion Rate in SN1987A}

Spherically symmetric accretion by a compact star in an initially static, infinite, background was described by
\citet{Bondi:1952kx} (see also, e.g.,  \citealt{Shapiro:1983fj}).
The mass accretion rate $\dot{M}$ is obtained from the density and sound velocity at infinity,
$\rho_\infty$ and $c_\infty$ respectively, as
\begin{equation}
\dot{M_\mathrm{B}} = 4 \pi \lambda  \left(\frac{GM}{c_\infty^2}\right)^2 \rho_\infty c_\infty,
\label{Eq:Bondi}
\end{equation}
where the numerical constant $\lambda =1/\sqrt{2} \simeq 0.707$ for the case of an ideal gas with adiabatic index $\gamma =4/3$.
In our case the medium is not strictly initially at rest but has been set into expansion by the supernova shock wave.
At early times the core is in homologous expansion, with a velocity $v = r/t$ and density $\rho$ such that $\rho t^3 = \rho_a t_a^3$ is constant,
in terms of a reference density $\rho_a$ at  time $t_a$.
As long as the time $t$ is smaller than the Bondi accretion time scale
$\tau_\mathrm{B} \simeq GM/c_\infty^3$, one can still estimate $\dot{M}$ with Eq. (\ref{Eq:Bondi})
by allowing $\rho_\infty$ and $c_\infty$ to be time dependent, and obtain \citep{Chevalier:1989jk}
\begin{equation}
\dot{M_\mathrm{B}} = 5.77  \frac{(GM)^2}{K^{3/2}} (\rho_a t_a^3)^{1/2} \, t^{-3/2}.
\label{Eq:Bondi2}
\end{equation}

According to \citet{Woosley:1988yg} and \citet{Shigeyama:1988yq}, $\rho_a t^3_a \sim 10^9$ g cm$^{-3}$ s$^3$. Now the density and sound velocity at infinity have been estimated, for SN1987A, by 
\citet{Woosley:1988yg} and \citet{Bethe:1990zl} as being of the order of
\begin{eqnarray}
\rho _{\infty }   &=&   \frac{2.5\,\mathrm{M}_{\odot }}{(4/3)\pi (v_{f}t)^{3}}
                     \simeq 1.78\times 10^{-13}\left( \frac{t}{\mathrm{yr}}\right) ^{-3} \,
                                 \mathrm{g}\,\mathrm{cm}^{-3}, 
\\
c_{\infty }^{2}   &=&   \frac{\gamma k_{B}T_{\infty }}{\mu m_{H}}
                    \simeq  1.24\times 10^{12}\left( \frac{t}{\mathrm{yr}}\right) ^{-\frac{3}{4}} \,
                                 \mathrm{cm}^{2}\,\mathrm{s}^{-2},
\end{eqnarray}
where $v_{f}\simeq 600\,\mathrm{km}\,\mathrm{s}^{-1}$ is the final expansion velocity and
$T_{\infty } \simeq 70\,\mathrm{keV} ( 4\times 10^{9}\,\mathrm{cm}/v_{f}t) ^{3/4}$
for a radiation dominated shock as in a supernova like SN1987A.
\citet{Bethe:1990zl} calculated that the temperature for a shock radius of $4\times 10^{9}\,\mathrm{cm}$ is 
$T_{sh}\simeq 70\,\mathrm{keV}$.
The mass of the expanding CO core is $\sim 4\,\mathrm{M}_{\odot }$, but here we considered only $2.5\,\mathrm{M}_{\odot }$ because $1.5\,\mathrm{M}_{\odot }$ were taken to make the compact object at the center of the supernova. We hence have
\begin{equation}
\dot{M} \simeq 2.23 \times 10^{22} \left( \frac{t}{\mathrm{yr}}\right) ^{-\frac{15}{8}} \,
                        \mathrm{g} \,\mathrm{s}^{-1}
              \simeq 3.5 \times10^{-4} \left( \frac{t}{\mathrm{yr}}\right)^{-\frac{15}{8}} \,
                         \mathrm{M}_{\odot } \,\mathrm{yr}^{-1}.
\end{equation}
\citet{Woosley:1988yg} calculated the time that the reverse shock takes to return to
the surface of the neutron star for SN1987A as
$t  \simeq 7 \times 10^3$ s, 
giving us, for the accretion rate in SN1987A in the hypercritical regime
\begin{equation}
\dot{M}  \simeq 1.57\times 10^{29} \,  \mathrm{g}\,\mathrm{s}^{-1}=2500 \; \mathrm{M}_{\odot} \,\mathrm{yr}^{-1}.
\end{equation}
This accretion rate exceeds by an order of magnitude the value calculated by \citet{Chevalier:1989jk}, 
$\dot{M}=2.2 \times 10^{28}$~g~s$^{-1}=340$~M$_{\odot}$~yr$^{-1}$ because of different assumed values at infinity. 
Now the Eddington mass accretion rate when considering photon radiation is 
$ \dot{M}_{Edd}=3.77\times 10^{18}\,\mathrm{g}\,\mathrm{s}^{-1}$, if one considers electron scattering in pure ionized Hydrogen as the source of opacity,  $k_{es} =  0.4\,\mathrm{cm}^{2}\,\mathrm{g}^{-1}$. When $\dot{M}>>\dot{M}%
_{Edd} $\ the flow is what we described above as \textit{Hypercritical Flow}, studied by \citet{Blondin:1986sf}. 
For the case of SN1987A, we have
\begin{equation}
\frac{\dot{M}}{\dot{M}_{Edd}}\simeq 10^{9}-10^{10},
\end{equation}
for the two values given above, placing such flows clearly in the hypercritical, neutrino cooled regime. 
Henceforth we adopt as our fiducial accretion rate the value 
\begin{equation}
\dot{M}_0  \simeq 2.2 \times 10^{28} \,  \mathrm{g}\,\mathrm{s}^{-1} = 340 \; \mathrm{M}_{\odot} \,\mathrm{yr}^{-1}.
\label{Equ:M_0}
\end{equation}
%

\subsection{The Envelope and the Shock Radius: Spherical Case}

When the reverse shock bounces against the surface of the neutron star, a third
expansive shock is formed, which propagates through the infalling matter.
Thus, eventually an atmosphere in quasi--hydrostatic equilibrium is formed around the compact object
(see Panel D in Figure~\ref{Fig:scenario}), 
whose general structure can be calculated analytically under some simplifying assumptions.

Following the formulation of \citet{Chevalier:1989jk},
the structure of the envelope
is calculated and an expression for the pressure at the surface of the
neutron star, $P_{ns}$, in terms of $\dot{M}$ and the shock
radius, $r_{sh},$ is derived. Cooling by neutrinos close to the neutron
star surface, which depends on  $P_{ns}$, is introduced and this
ultimately determines the shock radius solely as a function of the accretion rate. 
From the condition 
of hydrostatic equilibrium, $dP/dr=-\rho GM/r^{2},$ we obtain the integrated
values for the pressure, density and velocity as a funtion of the distance
from the neutron star surface. This is possible because neutrino cooling is only
important near the surface of the neutron star and we can consider that the post--shock
flow is adiabatic over the  greater part of the volume. In addition, the flow is
highly subsonic except close the shock. With this we obtain, $P\propto r^{-\gamma /\left(
\gamma -1\right) }\propto r^{-4},\rho \propto r^{-1/\left( \gamma -1\right)
}\propto r^{-3},v\propto r^{\left( 3-2\gamma \right) /\left( \gamma
-1\right) }\propto r$, where we have used $\gamma =4/3$. 
These results also are valid in the shock and the envelope structure in hydrostatic
equilibrium is,
\begin{eqnarray}
P &=&P_{sh}\left( \frac{r}{r_{sh}}\right) ^{-4}, \\
\rho &=&\rho _{sh}\left( \frac{r}{r_{sh}}\right) ^{-3}, \\
v &=&v_{sh}\left( \frac{r}{r_{sh}}\right) ,
\end{eqnarray}
where $P_{sh},$ $\rho _{sh},$ $v_{sh}$ and $r_{sh}$ are the values at the shock. 
These values can be obtained from the jump conditions when $P_{sh}>>P_{0}$, where 
$P_{0}$ and $\rho _{0}$\ the pre--shock pressure and density. Under these
considerations we obtain $P_{sh}=(7/8)\rho _{0}v_{0}^{2},$ $\rho _{sh}=7\rho
_{0}$ and $v_{sh}=-(1/7)v_{0}$. 
The pre--shock velocity is that of free-fall, $v_{0}=\sqrt{2GM/r_{sh}}$, and
the density is $\rho _{0}=\dot{M}/4\pi r_{sh}^{2}v_{0}$.
 From equation (8) we obtain the pressure at the surface,
$P_{ns}=1.36\times 10^{-12}\dot{M}r_{sh}^{3/2}$, with $M\sim
1.44\,\mathrm{M}_{\odot }$ and $r_{ns}\sim 10^{6}\,\mathrm{cm}$. 
On the other hand, the energy loss by neutrinos (only pair production) by 
unit volume can be estimated as \citep{Dicus:1972eu},
\begin{equation}
\dot{\varepsilon }_{n}=1.83\times 10^{-34}P^{2.25}\,\mathrm{erg}%
\,\mathrm{cm}^{-3}\,\mathrm{s}^{-1}.
\end{equation}
In this case, we consider that e$^{\pm}$ pairs also contribute to the pressure. Now, this cooling is operative only in a small volume
close to the neutron star surface, $\sim \pi r_{ns}^{3}$ since it is a sensitive function of temperature. So, from energy conservation, the shock radius is obtained as,
\begin{equation}
r_{sh}\simeq 7.58\times 10^{8}\left( \frac{\dot{M}}{\,\mathrm{M}%
_{\odot }\,\mathrm{yr}^{-1}}\right) ^{-\frac{10}{27}}\,\mathrm{cm}.
\end{equation}

With this we have the structure of the envelope and the shock radius as a function
of the accretion rate. For the case of SN1987A with our fiducial accretion rate  $\dot{M}=\dot{M}_0=340$~$\mathrm{M}_{\odot }$~yr$^{-1}$, the shock radius is $r_{sh}\simeq 8.77\times 10^{7}\,\mathrm{cm.}$

\subsection{The Envelope and the Shock Radius: The Accretion Column}

\begin{figure}[!t]\centering
  \includegraphics[width=0.40\columnwidth]{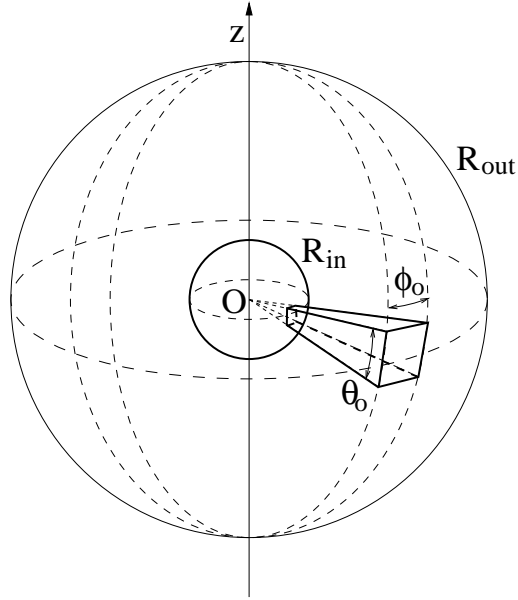}
  \caption{Schematic geometry of the accretion column onto the neutron star surface.}
  \label{Fig:schema}
\end{figure}

If we consider a small rectangular accretion column of area $A_{\rm col}$ onto a fraction of the neutron star surface, 
with area $A=4 \pi R_{\rm NS}^{2}$, we can take it to be
a plane--parallel surface (see Figure~\ref{Fig:schema}).
In this case, the spherical mass accretion rate must be scaled to its
value in the column. Since in the spherical case the area depends on the
distance to the neutron star, while for the case of an accretion column the area is constant,
the structure of the envelope and the shock radius are modified. 
Note that this modification causes
the mass accretion rate per unit area to be independent of height in the domain.
Also, since it is smaller than in the spherical case, the analytical estimate of the shock radius decreases significantly, and is now given by
\begin{equation}
y_{sh}\simeq 7.61\times 10^{6}\left( \frac{\dot{M}/A}{\,\mathrm{M}%
_{\odot }\,\mathrm{yr}^{-1}/A_{\rm col}}\right) ^{-0.16}\,\mathrm{cm},
\label{eq:shcol}
\end{equation}
where $y$  measures the height above the neutron star surface. With these considerations and taking $A_{\rm col}=(3\times 10^{5})^{2}$~cm$^{2}$, the shock radius for the fiducial accretion rate, Eq. (\ref{Equ:M_0}), is $y_{sh}=6.92\times 10^{6}\, \mathrm{cm}$, and the structure of the envelope is,
\begin{eqnarray}
P &=&P_{sh}\left( \frac{y}{y_{sh}}\right) ^{-4}, \\
\rho  &=&\rho _{sh}\left( \frac{y}{y_{sh}}\right) ^{-3}, \\
v &=&v_{sh}\left( \frac{y}{y_{sh}}\right) ^{3}.
\end{eqnarray}

The velocity profile is different for the accretion column as well because the column area is constant as a function of height above the neutron star. 
The conditions in the shock in the SN1987A scenario are thus 
\begin{eqnarray}
v_{0} &=&\sqrt{\frac{2GM}{y_{sh}}}\simeq 7.53\times 10^{9}\,\mathrm{cm}\,%
\mathrm{s}^{-1}, \\
\rho _{0} &=&\frac{\dot{M}}{v_{0}\times \rm A}\simeq 2.31\times 10^{5}\,\mathrm{g}\,\mathrm{cm}^{-3}, \\
\rho _{sh} &=&7\rho _{0}\simeq 1.62\times 10^{6}\,\mathrm{g}\,\mathrm{cm}%
^{-3}, \\
v_{sh} &=&-\frac{1}{7}v_{0}\simeq -1.07\times 10^{9}\,\mathrm{cm}\,\mathrm{s}%
^{-1}, \\
P_{sh} &=&\frac{7}{8}\rho _{0}v_{0}^{2}\simeq 1.14\times 10^{25}\,\mathrm{dyn%
}\,\mathrm{cm}^{-2}.
\end{eqnarray}

Now we can build a numerical model with more refined physics to perform
2D hydrodynamics (HD) and magnetohydrodynamics (MHD) simulations, which we can compare with the 1D analytical results. 

\section{Numerical Approach}
\label{sec:numerical}

For the work shown in this paper we used the numerical code AMR FLASH2.5 \citep{Fryxell:2000yg}
 to perform the 2D simulations. FLASH (http://flash.uchicago.edu/website/home/) is a modular, portable, highly scalable,
adaptive-mesh simulation code for astrophysical hydrodynamics problems. It
was originally developed at the DOE ASCI Alliances Center for Astrophysical
Thermonuclear Flashes at the University of Chicago for the purpose of
simulating Type Ia supernovae, novae, and X-ray bursts. It has since evolved
to handle more general astrophysical problems, including those involving
collisionless particle dynamics. FLASH is freely available from the ASCI
Flash Center. This code is designed to allows users to configure initial and
boundary conditions, change algorithms, and add new physics modules with
minimal effort. It uses the PARAMESH library to manage a block-structured
adaptative grid, placing resolution elements where they are needed most.

\subsection{The Numerical Method}

FLASH2.5 provides two main types of modules: Physics and
Infrastructure Modules. In our model we used the hydro--mhd, eos--helmholtz,
gravity and neutrino--cooling custom modules. 

The FLASH code solves the the equations of a magnetized fluid (ideal or
non--ideal), described by
\begin{eqnarray}
\frac{\partial \rho }{\partial t}+\mathbf{\nabla }\cdot \left( \rho \mathbf{v%
}\right) &=&0, \\
\frac{\partial \rho \mathbf{v}}{\partial t}+\mathbf{\nabla }\cdot \left(
\rho \mathbf{vv-BB}\right) +\nabla P_{\ast } &=&\rho \mathbf{g+\nabla \cdot
\tau ,} \\
\frac{\partial \rho E}{\partial t}+\mathbf{\nabla }\cdot \left[ \mathbf{v}%
\left( \rho E+P_{\ast }\right) -\mathbf{B}\left( \mathbf{v}\cdot \mathbf{B}%
\right) \right] &=&\rho \mathbf{v\cdot g+}O(\mathbf{\tau },\eta ), \\
\mathbf{\nabla \cdot }\left( \mathbf{v}\cdot \mathbf{\tau +}\sigma \nabla
T\right) +\mathbf{\nabla }\cdot \left( \mathbf{B}\times \left( \eta \mathbf{%
\nabla }\times \mathbf{B}\right) \right) &=&O(\mathbf{\tau },\eta ), \\
\frac{\partial \mathbf{B}}{dt}+\mathbf{\nabla }\cdot \left( \mathbf{vB}-%
\mathbf{Bv}\right) &=&-\mathbf{\nabla }\times \left( \eta \mathbf{\nabla }%
\times \mathbf{B}\right),
\end{eqnarray}
where
\begin{eqnarray}
P_{\ast } &=&P+\frac{B^{2}}{2}, \\
E &=&\frac{1}{2}v^{2}+\varepsilon +\frac{B^{2}}{2\rho }, \\
\mathbf{\tau } &\mathbf{=}&\mu \left[ \left( \nabla \mathbf{v}\right)
+\left( \nabla \mathbf{v}\right) ^{\mathbf{T}}-\frac{2}{3}\left( \nabla 
\mathbf{v}\right) \right] .
\end{eqnarray}
Here $P_{\ast },$ $E$ and $\mathbf{\tau }$ are the total pressure, total specific
energy and stress tensor, respectively, and the remaining symbols have their usual meaning. 
Units in these equations are such that no $4\pi $ and $\mu _{0}$ factors appear. 

We have simplified the above set of equations by restricting ourselves to the ideal hydro and MHD cases. Setting the thermal conductivity, $\sigma$, and electrical resistivity, $\eta$, to zero is justified by the fact that time scales for
heat and magnetic field diffusion are many orders of magnitude larger than our simulation times.
The inviscid ($\mu=0$) approximation, i.e., neglect of momentum diffusion, is acceptable because we have not considered rotation in our models.
Note that when $\mathbf{B}=0,$ the Euler equations are then obtained.

A particular complication associated with solving the MHD equations
numerically lies in the solenoidality of the magnetic field. The non-existence of magnetic monopoles,  $\mathbf{\nabla }\cdot \mathbf{B=0}$ is difficult to satisfy in discrete computations. Being only an initial
condition of the MHD equations, it enters the equations indirectly and is
not, therefore, guaranteed to be generally satisfied unless special
algorithmic provisions are made. FLASH2.5 uses a simple yet very
effective method to destroy the magnetic
monopoles on the scale on which they are generated. In this method, a
diffusive operator proportional to $\nabla \nabla \cdot \mathbf{B}$ is added
to the induction equation, so that the equations become
\begin{equation}
\frac{\partial \mathbf{B}}{dt}+\mathbf{\nabla }\cdot \left( \mathbf{vB}-%
\mathbf{Bv}\right) =-\mathbf{\nabla }\times \left( \eta \mathbf{\nabla }%
\times \mathbf{B}\right) -\mathbf{v\nabla \cdot B+}\eta _{a}\nabla \nabla
\cdot \mathbf{B},
\end{equation}
with the artificial diffusion coefficient $\eta _{a}$\ chosen to mach that
of grid numerical diffusion. In the FLASH code, $\eta _{a}=(\lambda
/2)(1/\Delta x+1/\Delta y+1/\Delta z)^{-1},$ in 3D where $\lambda $\ is the
largest characteristic speed in the flow. Since the grid magnetic diffusion
Reynolds number is always on the order of unity, this operator locally
destroys magnetic monopoles at the rate which they are created.
All our simulations are in cartesian coordinates:
in the presence of a magnetic field polar/spherical coordinates are very troublesome and, presently, the MHD version of FLASH does not support them.

\subsection{The Physics Ingredients}

In the analytical approach an ideal gas equation of state has been used.
This allows much simplification in the structure of the envelope
and in addition, the flow  can be managed like an adiabatic fluid
with $\gamma =4/3$. The gas is dominated by radiation, which is
trapped within the flow. Also, the neutrino losses depend on a high power of
the pressure, but are only important at the base of the envelope. 
Nevertheless, to account for the thermodynamics more accurately and for the consequent piling up of matter on the star,
 it is advisable and necessary to work with a more complete and realistic equation of state. 
 The Helmholtz EOS provided with the FLASH2.5
distribution contains more physics and is appropiate for addressing
astrophysical phenomena in which electrons and positrons may be relativistic
and/or degenerate and in which radiation may significantly contribute to the
thermodynamic state. This EOS thus includes contributions from black-body radiation,
completely ionized ideal nuclei, and free electrons and positrons. The pressure and internal energy are calculated as the sum over
the components,
\begin{eqnarray}
P_{tot} &=&P_{rad}+P_{ion}+P_{ele}+P_{pos}+P_{coul} \\
\varepsilon _{tot} &=&\varepsilon _{rad}+\varepsilon _{ion}+\varepsilon
_{ele}+\varepsilon _{pos}+\varepsilon _{coul}.
\end{eqnarray}
Here the subscripts \textquotedblleft rad\textquotedblright ,
\textquotedblleft ion\textquotedblright , \textquotedblleft
ele\textquotedblright , \textquotedblleft pos\textquotedblright\ and
\textquotedblleft coul\textquotedblright\ represent the contribution from 
radiation, nuclei, electrons, positrons, and Coulomb corrections, respectively. The radiation portion assumes a blackbody in local
thermodynamic equilibrium, the ion portion (nuclei) is treated as an ideal
gas with $\gamma =5/3,$ and the electrons and positrons are treated as a
non--interacting Fermi gas of arbitrary degeneracy and relativity. 

Under the physical conditions of interest for the set of simulations presented here, the gas is dense enough that the optical depth for photons is $\tau_{\gamma}\gg 1$, and they are fully trapped in the flow. Adding the corresponding term to the pressure as $P_{rad}=aT^{4}/3$ is thus entirely appropriate. We note that more recent versions of FLASH (upwards of 3.2) include modules for radiation transport, making them useful for a wider range of studies. 

The gravity module suplied with FLASH2.5 computes gravitational source terms
for the code. These  can take the form of the gravitational
potential $\phi (\mathbf{x})$ or the gravitational acceleration,
\begin{equation}
g(\mathbf{x})=-\nabla \phi (\mathbf{x}).
\end{equation}
The gravitational field can be externally imposed or self--consistently
computed from the gas density via the Poisson equation,
\begin{equation}
\nabla ^{2}\phi (\mathbf{x})=4\pi G\rho (\mathbf{x}),
\end{equation}
where $G$ is Newton's gravitational constant. In the latter case, either
periodic or isolated boundary conditions can be applied. In our case, we
used an externally applied gravitational field (\textit{plane--parallel
gravitational field}), where the acceleration vector is parallel to one of
the coordinate axes, and its magnitude drops with distance along that
axis as the distance squared. Its magnitude and direction are
independent of the other two coordinates.

In the conditions present in both the high density part of the accretion flow 
and the underlying envelope neutrino emission occurs essentially through neutral currents processes.
The five processes we included in the models are analogous to standard photon emission 
processes where the $\gamma$ emission is replaced by a $\nu - \overline{\nu}$ pair.
They are:

{PAIR ANNIHILATION:}  $e^{-}+e^{+}\rightarrow \nu +\overline{\nu }$,

{PHOTONEUTRINOS:}  $\gamma +e^{\pm }\rightarrow e^{\pm }+\nu +\overline{\nu }$,
the analogous of Compton scattering,

{PLASMON DECAY:} $\Gamma \rightarrow \nu +\overline{\nu }$,
where $\Gamma$ is a plasmon,

{BREMSSTRAHLUNG:} $e^{\pm }+N\rightarrow e^{\pm }+N+\nu +\overline{\nu }$,
where $N$ is a nucleus, and

{SYNCHROTRON:} $e^{\pm }+\mathbf{B}\rightarrow e^{\pm }+\mathbf{B}+\nu +\overline{\nu }$,
where $\mathbf{B}$ represents the magnetic field.

For the first four processes we used the calculations of  \citet{Itoh:1996fv} 
and for the synchrotron emission we followed \citet{Bezchastnov:1997jk}.
Pair annihilation is the dominant process but synchrotron can make some significant contribution
when the magnetic field becomes strongly compressed. As noted above, the density in the flow is typically high enough that photons are trapped, but not neutrinos. As a rough guide, the optical depth for neutrinos under coherent scattering off free nuclei is $\tau_{\nu}\simeq 1$ when $\rho\simeq 10^{11}$~g~cm$^{-3}$, which is several orders of magnitude higher than the maximum values studied here. Thus neutrino cooling can be implemented simply as a sink in the energy equation.

\subsection{The Initial and Boundary Conditions}

We simulated a small 2D accretion column in cartesian coordinates anchored onto the surface of the neutron star, and considered various accretion rates and magnetic field configurations. This set of simulations allows us to compare numerically obtained results in the pure hydrodynamical and MHD case with the proposed
analytical approach, as well as to analyze the reaction of the magnetic field
to the infalling gas. 
The computational domain covers $0 \leq x \leq 3 \times10^{5}$~cm, $0 \leq y \leq 10^{7}$~cm. 
The dimensions for the column are: $\rm A_{\rm col}%
=L_{x}\times L_{z}=(3\times 10^{5})^{2}\,\mathrm{cm}^{2}$ (for the base) and $\rm 
height_{col}=L_{y}=10^{7}\,\mathrm{cm}$ for the height. This height is adequate because it is below
the analytical shock radius value calculated for the accretion rate of
SN1987A. The fluid is initially in free fall and we set a constant temperature in the gas. 
We considered horizontal ($B_{x}=10^{12}\,\mathrm{G}$, $B_{y}=0$),
vertical ($B_{x}=0$, $B_{y}=10^{12}\,\mathrm{G}$, mimicking accretion onto the magnetic pole of the neutron star), diagonal ($B_{x}=B_{y}=10^{12}\,%
\mathrm{G}$) and dipolar ($B_{x}=2\mu /y^{3}$, $B_{y}=0$, representing accretion onto the neutron star equator) cases, where $\mu =5\times 10^{29}$
is the dipolar moment, fixed so that $B_{x}=10^{12}\,\mathrm{G}$ at
the neutron star surface. With these considerations, the initial conditions in the
column for velocity, temperature and density are:
\begin{eqnarray}
\rho  &=&\frac{\dot{M}_{\rm col}}{v_{ff}\times\rm A_{\rm col}}, \\
T &=&10^{9}\,\mathrm{K}, \\
v_{ff} &=&\sqrt{\frac{2GM}{y}}, \\
B &=&10^{12}\,\mathrm{G},
\end{eqnarray}
where $\dot{M}_{\rm col}=\dot{M}A_{\rm col}/A$ is the scaled accretion rate in the column making up the domain. 

For the vertical boundaries of the accretion column, $x=0$ and $x=3\times 10^{5}$~cm, parallel to the $y$-axis, we implement standard periodic boundary conditions. Thus, any fluid element moving out of the computational domain on the right (left) boundary re-enters the domain on the left(right) edge with the same thermodynamical properties and velocity.

For the top and bottom of the computational domain, parallel to the $x$-axis, we implemented custom boundary conditions. The gravity vector is along the $y$-axis, and we want the lower boundary at $y=0$ to support the fluid above against infall, mimicking the hard surface of the neutron star, and in addition to have the magnetic field anchored to it. In order to establish this
boundary, we use ``guard", or ``ghost" numerical cells. These are cells outside the formal computational domain (e.g., at $y\leq 0$ or $y\geq 10^{7}$) for which we can fix the hydrodynamical and thermodynamical properties and that are not evolved along with the rest of the flow. They are useful precisely to guarantee boundary conditions of interest, depending on the setup of the problem. A layer of at least 2 such cells along the top and bottom of the domain can thus be used to compute proper gradients at the edge of the flow (e.g., a pressure or temperature gradient). In this case, along the bottom edge of the column, we fix the velocities to be null in all guard cells, $(v_{x}=v_{y}=v_{z}=0)$, keeping them at rest, and copy the
density and the pressure of the first zone of the numerical domain to mimick 
the the neutron star surface: $\left( \rho =\rho(1),P=P(1)+\rho
v^{2}+\rho gh\right)$, where the label $1$ refers to the first cell in the computational domain. The
magnetic field is put in this boundary in such form that it is continuous
from the guard cell to the physical domain, i.e, we anchor the magnetic
field onto the neutron star surface and in the rest of the guard cells it is null.
The other thermodynamics variables are calculated from the equation of state. 
At the top of the column, $y=10^{7}$~cm, we set the velocity to be that of free fall, $v_{y}=-\sqrt{2GM/y}$ in all the
guard cells, and set the density to fix a constant inflow mass accretion rate,
 $\rho =\dot{M}_{\rm col}/(|v_{y}|\times\rm A_{\rm col})$. 
As in the computational domain initially, the temperature in the guard cells is set to $T=10^{9}$~K (at all times).
The remaining variables are calculated from the equation of state.

\section{Results and Discussion}
\label{sec:results}

We now present results obtained from the 2D hydrodynamical simulations (HYDRO) as well as for the MHD case for an accretion column in cartesian coordinates, and compare these to the analytical scalings. We varied the accretion rate and magnetic field configuration (for the MHD case). The chosen rates were one, two and three orders of magnitude above our fiducial rate $\dot{M}_{0}=340\, \mathrm{M}_{\odot }\,\mathrm{yr}^{-1}$. 

\subsection{Comparison of the HYDRO and MHD solvers }

For the assumed physical parameters of SN1987A, in $200\,\mathrm{ms}$ the system
reaches a quasi--stationary state, whereas for higher rates of
accretion, this drops substantially: 60~ms at $10\dot{M}_{0}$, 20~ms at $100\dot{M}_{0}$ and 5~ms at $1000 \dot{M}_{0}$. We set a level of refinement of 4, with 
2 blocks along the $x$-axis and 18 along $y$-axis. This implies an effective resolution of $128\times 1152$ zones in the computational domain. 

\begin{figure}[!t]\centering
  \includegraphics[width=0.60\columnwidth]{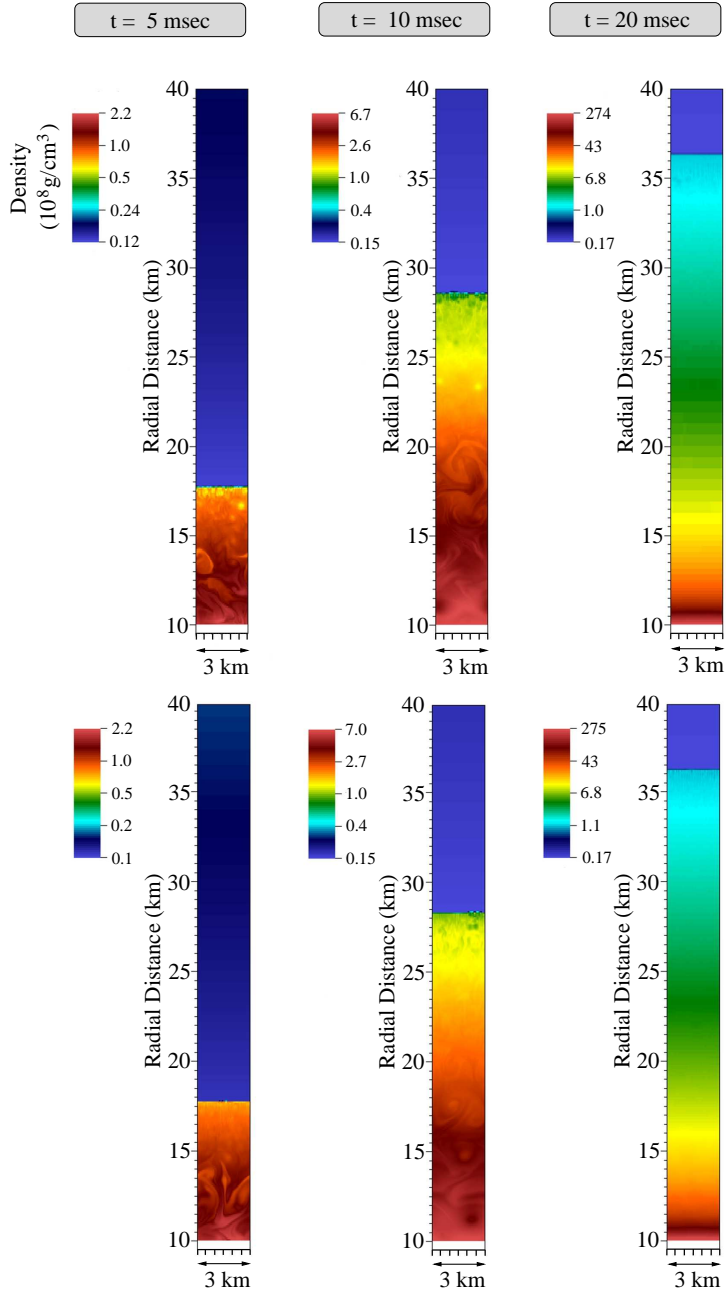}
  \caption{Color maps of density for cases HYDRO (top) and MHD\_0 (bottom) at $t=5, 10, 20$~ms from left to right. The accretion rate is $\dot{M}=100\dot{M}_{0}$.}
  \label{Fig:maps1a}
\end{figure}

\begin{figure}[!t]\centering
  \includegraphics[width=0.80\columnwidth]{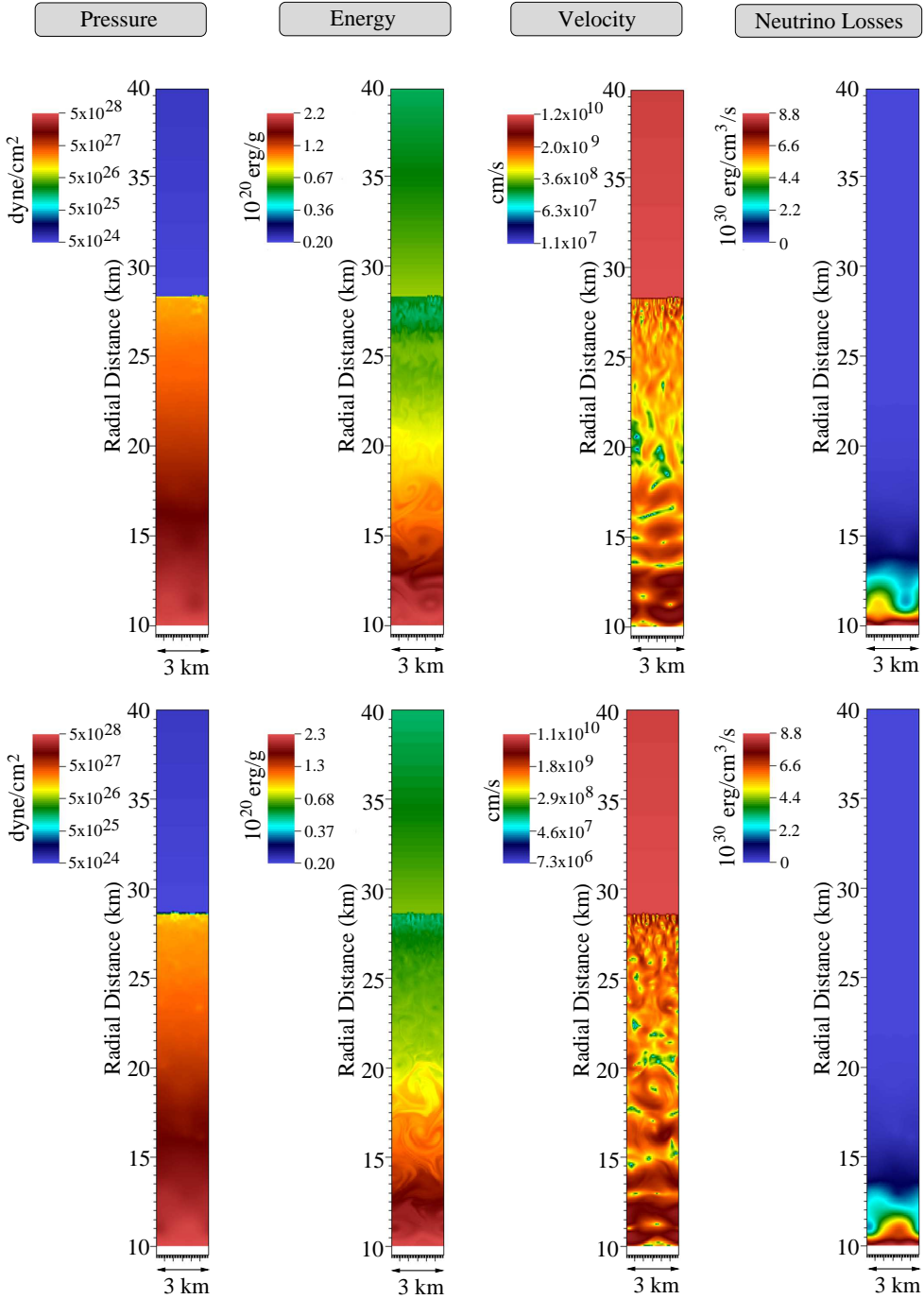}
  \caption{Color maps of pressure, total specific energy, magnitude of velocity and neutrino emissivity, for cases HYDRO (top) and MHD\_0 (bottom) at $t=10$~ms. The accretion rate is $\dot{M}=100\dot{M}_{0}$.}
  \label{Fig:maps1b}
\end{figure}

\begin{figure}[!t]\centering
  \includegraphics[width=0.50\columnwidth]{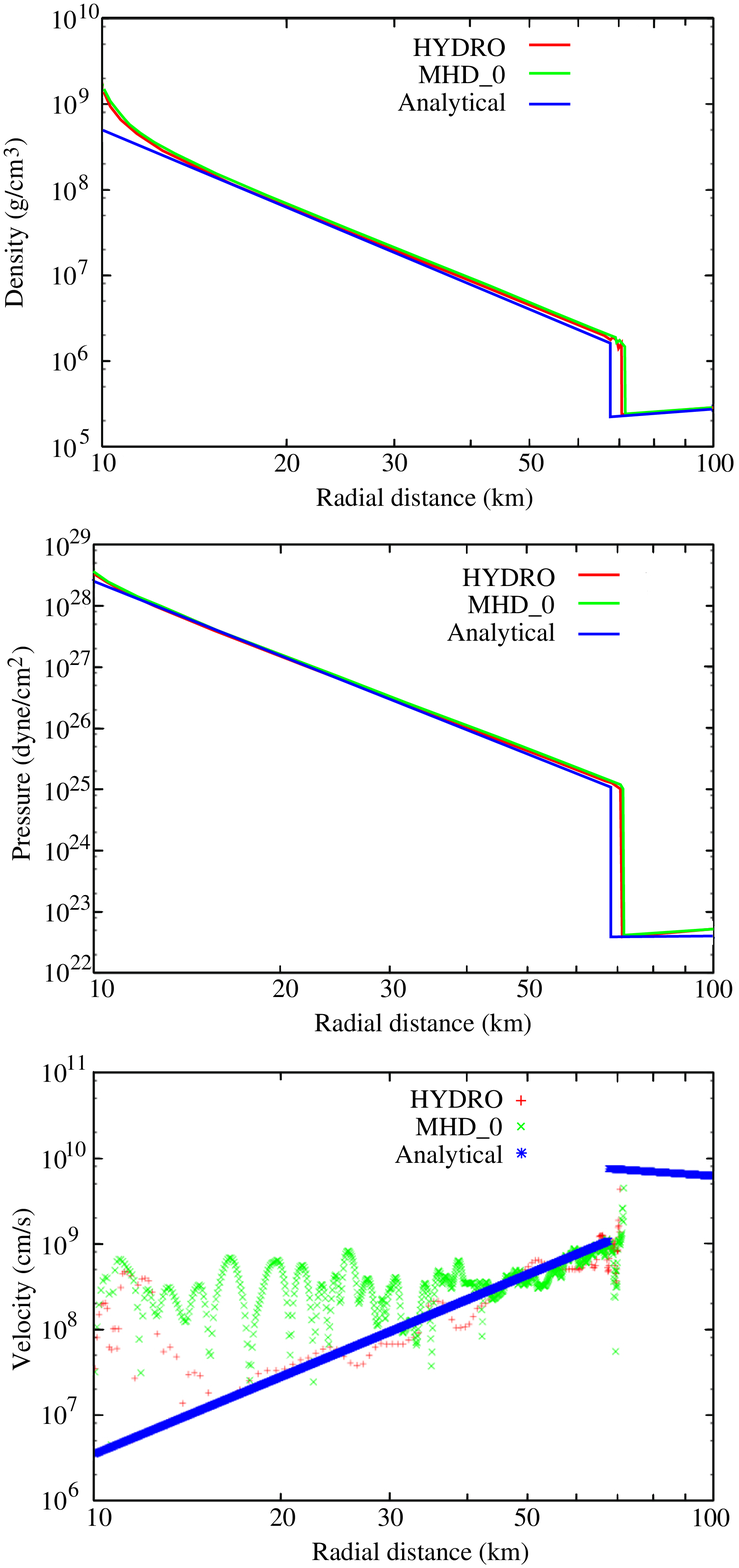}
  \caption{Radial profiles of density (top), pressure (middle) and magnitude of velocity (bottom) at $t=200$~ms. The accretion rate is $\dot{M}=\dot{M}_{0}$. The HYDRO (red), MHD\_0 (green) and analytical solution (blue) are shown together. The location of the shock is reproduced with very good agreement in the two numerical cases with respect to the analytical solution. In the shocked region, the velocity shows significantly higher behavior due in part to lateral motions of the gas. At small radii, the numerical solutions deviate from the analytical curve, since the assumption of self--similarity is no longer valid as material piles up near the star and cooling becomes relevant in the energy balance.}
  \label{Fig:prof1a}
\end{figure}

In Fig. \ref{Fig:maps1a} we show the density contrast for the HYDRO and MHD cases with null magnetic field (MHD\_0), for $\dot{M}=100\dot{M}_{0}$. We choose
this accretion rate as being  representative since its associated shock radius
is much smaller than for $\dot{M}=\dot{M}_{0}$, and is therefore easier to
visualize. In addition, it is possible to both do a comparative analysis of
solvers (HYDRO and MHD) and of their response  to the imposed initial
conditions. 
This comparison allows us to determine whether the equations are being solved in both modules to a comparable accuracy. 
In principle, the  MHD module with null magnetic field should
reproduce exactly the results obtained with module HYDRO. 
The constrasts of
pressure, specific total energy, velocity and neutrino cooling per unit volume for all the cases (at $t=10\,\mathrm{ms}$), are shown in Fig.  \ref{Fig:maps1b}.
The radial profiles of density, pressure and velocity for the SN1987A accretion rate
are given in Fig. \ref{Fig:prof1a}. 

We note that although the system reaches a quasi--stationary state in $t=20\,\mathrm{ms}$, there is remnant noise in the radial profile of the velocity due to the interaction of the matter with the lower boundary condition and to the fact that horizontal motions are allowed because of the periodic boundary condition.
On the other hand, only the bottom section, $4\times 10^{6}\,\mathrm{cm}$, of the entire accretion column, with height $10^{7}\,\mathrm{cm}$ is shown, where the most interesting processes occur. We note that the
profiles, while not identical in all respects, are indeed very similar, showing that the HYDRO and MHD solvers are  giving essentially the same final state, both in space and time evolution. 
There is some convection in the early stages of the evolution, and Rayleigh-Taylor instabilities are present, but are quickly damped as the system approaches the stationary solution. Deviations from this are most evident when one examines variations in the velocity field. 

\begin{figure}[!t]\centering
  \includegraphics[width=0.60\columnwidth]{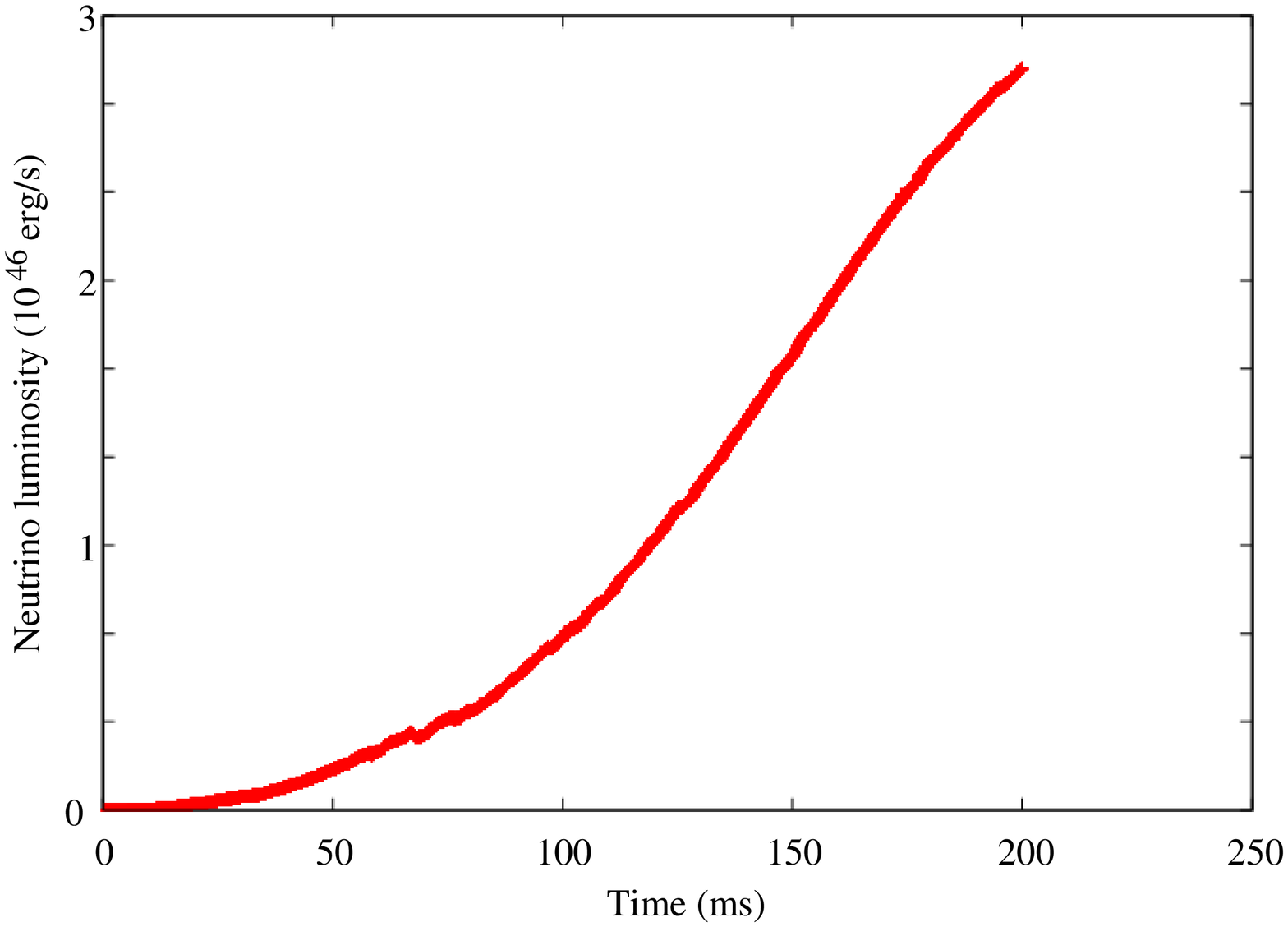}
  \caption{Neutrino luminosity integrated over the computational domain for the fiducial accretion rate, $\dot{M}=\dot{M}_{0}$, up to  $t=200\,\mathrm{ms}$. Note how after an initial transient, the power output is leveling off as the system reaches a quasi--stationary state.}
  \label{Fig:neutrino}
\end{figure}

\begin{figure}[!t]\centering
  \includegraphics[width=0.50\columnwidth]{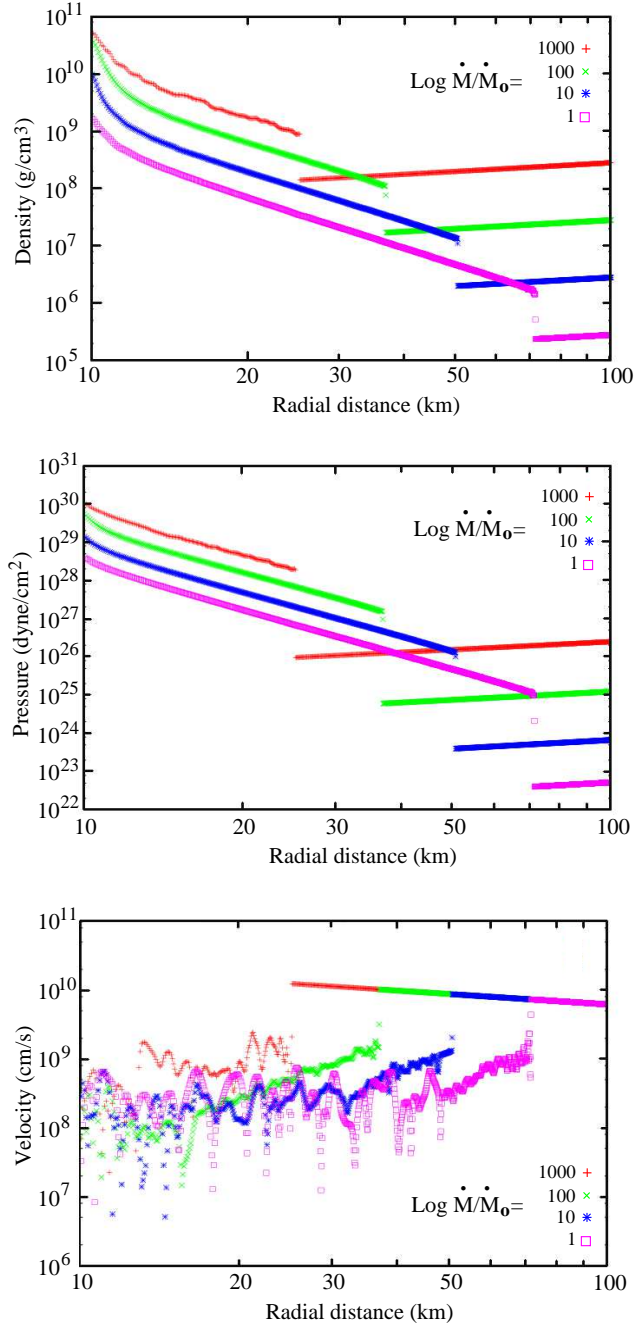}
  \caption{Radial profiles of density (top), pressure (middle) and velocity (bottom) for the MHD\_0 case after a stationary state has been reached. The accretion rates are $\dot{M}/\dot{M}_{0}= 1, 10, 100, 1000$ (cyan, blue, green and red, respectively).}
  \label{Fig:profs1b}
\end{figure}

\begin{figure}[!t]\centering
  \includegraphics[width=0.50\columnwidth]{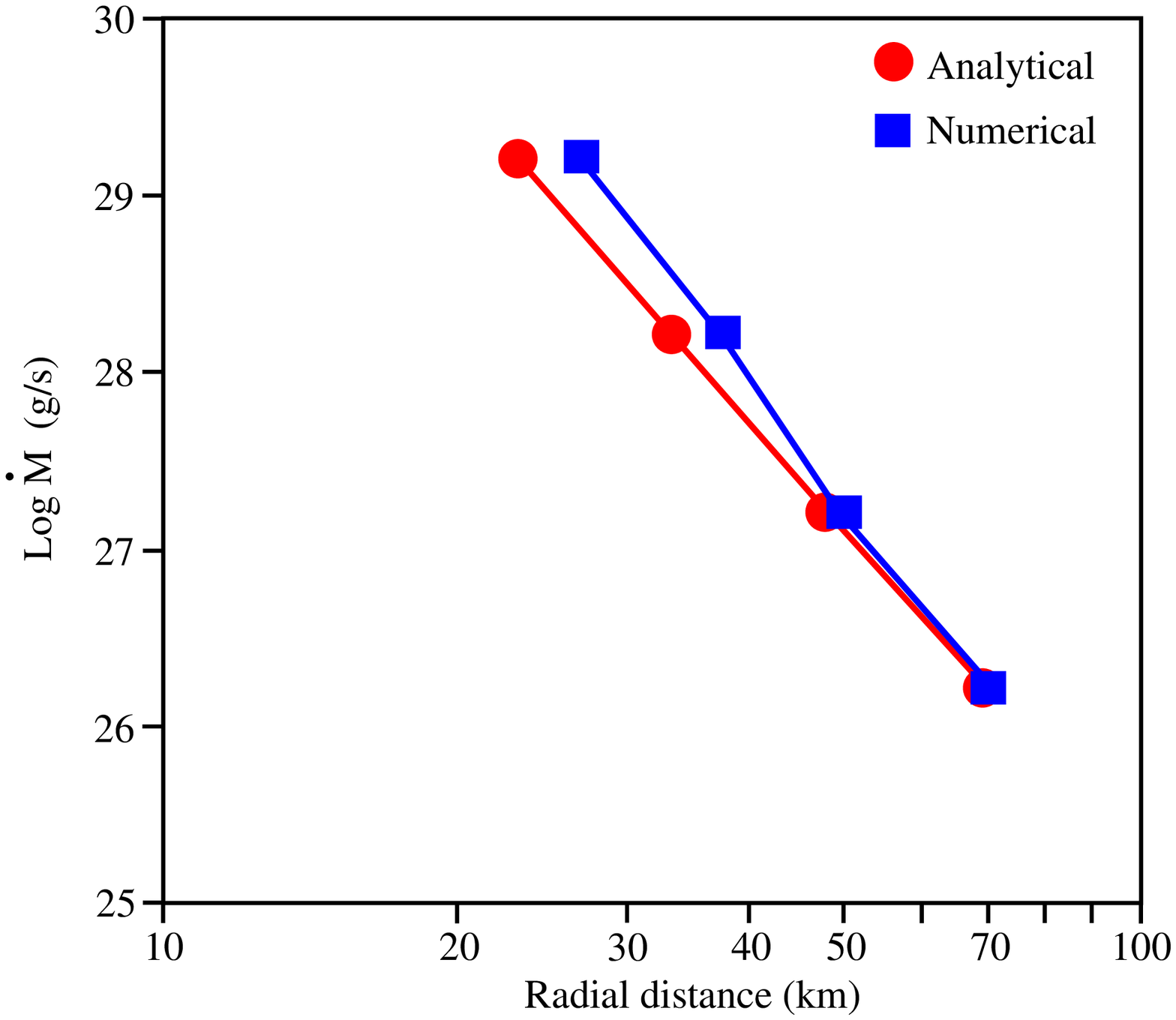}
  \caption{Accretion rate as a function of shock radius for the hydrodynamical simulations with $\dot{M}/\dot{M}_{0}=1000, 100, 10, 1$ after a stationary state has been reached. The analytical solution given by equation~\ref{eq:shcol} is shown for comparison.}
  \label{Fig:shockradius}
\end{figure}

The location of the shock is reproduced quite well, to within 5\% when compared to the analytical calculation. Moreover, both solvers place it essentially at the same height, indicating that the quantitative aspects are not affected from one to the other. Since the code is able to model the bottom of the column self--consistently within the imposed boundary condition, the numerical solution deviates from the self--similar scaling once neutrino cooling becomes important, and matter starts piling up near the surface. 

Hereafter, unless otherwise noted we refer to calculations with $\dot{M}=\dot{M}_{0}$. The adiabatic and radiative gradients can be calculated from the simulations, when
the system is relaxed. We find $\nabla _{ad}=1-1/\gamma _{c}\simeq 0.26,$ \ $%
\nabla _{rad}=\left( d\ln T/d\ln P\right) \simeq 0.24$. In this case, the
value of the adiabatic index $\gamma _{c}$ has been taken directly from the
simulation ($\gamma _{c}=1.35$), and the radiative gradient was calculated
by building a plot of temperature vs. pressure. These gradients have almost constant
values within the envelope, except in the region close to the neutron star surface. 
Since $\nabla _{ad}>\nabla _{rad},$ the system is manifestly stable to convection. Nevertheless, being so close numerically is probably indicative of 
marginal stability.  Within the envelope the flow is fully subsonic, as expected after passing through the accretion shock front: the sound speed is $c_{s}=\sqrt{\gamma
_{c}P/\rho }\simeq 6.9\times 10^{9}\,\mathrm{cm}\,\mathrm{s}^{-1}$, and  $v\simeq 1.24\times 10^{7}\,\mathrm{cm}\,\mathrm{s}^{-1}$, giving a Mach number  $m=v/c_{s}\simeq 10^{-3}$.
Therefore, besides confirming that the HYDRO and MHD solvers give accurate and consistent results, we are able to study the global structure of the accretion column in detail and compare it with the analytical approach, particularly in the region where the approximations in the latter break down. 

It is worthy to note the thermodynamical conditions the fluid is in as it accretes towards the proto--neutron star. The Fermi temperature can be computed from the Fermi energy $E_{F}$
\begin{equation}
T_{F}=\frac{E_{F}}{k_{B}}=\frac{\sqrt{p_{F}^{2}c^{2}+m_{e}^{2}c^{4}}%
-m_{e}c^{2}}{k_{B}}\simeq 6.48\times 10^{10}\,\mathrm{K,}
\end{equation}
at the base of the flow, where $p_{F}=(3\pi ^{2}n_{e})^{1/3}\hbar $ is the Fermi momentum.
The temperature obtained from the simulation, close to the bottom of the accretion column 
in quasi--stationary state is $T\simeq 4.54\times 10^{10}\,\mathrm{K}$, so $
T/T_{F}\simeq 0.7$. It is thus clear that assuming that the $e^{\pm}$ pairs are entirely degenerate is not a proper approximation, and a full expression such as the one in the Helmholtz equation of state is required if one wishes to compute the evolution of the flow accurately. 
It is also clear that neutrino cooling effectively turns on at a scale height $L_{y}\sim 2\times 10^{5}\,\mathrm{cm}$.  For the simulation
with $\dot{M}=\dot{M}_{0}$,  the integrated neutrino luminosity, shown in Fig. \ref{Fig:neutrino}, is $L_{\nu }\simeq 2.51\times 10^{46}\,\mathrm{erg}\,%
\mathrm{s}^{-1}$, close to the value estimated with the cooling function of 
 \citet{Dicus:1972eu} scaled to the column: $L_{\nu }=\dot{\varepsilon }_{n}\times
V\simeq 1.83\times 10^{46}\,\mathrm{erg}\,\mathrm{s}^{-1},$ with $%
V\simeq (2\times 10^{5})\times (3\times 10^{5})^{2}\,\mathrm{cm}^{3}.$

Once the system reaches the quasi--stationary state, radial profiles can be compared for different accretion rates. Four different rates for each initial condition were computed. In all of these, the piling up of material close to the neutron star surface is seen. The velocity profiles
remain noisy and turbulent in the shocked region, but on average the analytical profile is globally recovered. In Fig. \ref{Fig:profs1b} these are plotted, along with density and pressure, for case MHD\_0. Note also that at greater accretion rates the shock is located at lower height, as expected. 
For  $\dot{M}/\dot{M}_{0}=1, 10, 100, 1000$, the position of the shock in the simulation is at 
$R_{sh}/10^{6} {\rm cm}=7.06, 4.96, 3.74, 2.68$, in excellent agreement with the analytical values given by  $R_{sh}/10^{6} {\rm cm}=6.92, 4.80, 3.33, 2.31$, respectively (see Fig. \ref{Fig:shockradius}).

\subsection{Magnetic field submergence}

\begin{figure}[!t]\centering
  \includegraphics[width=0.50\columnwidth]{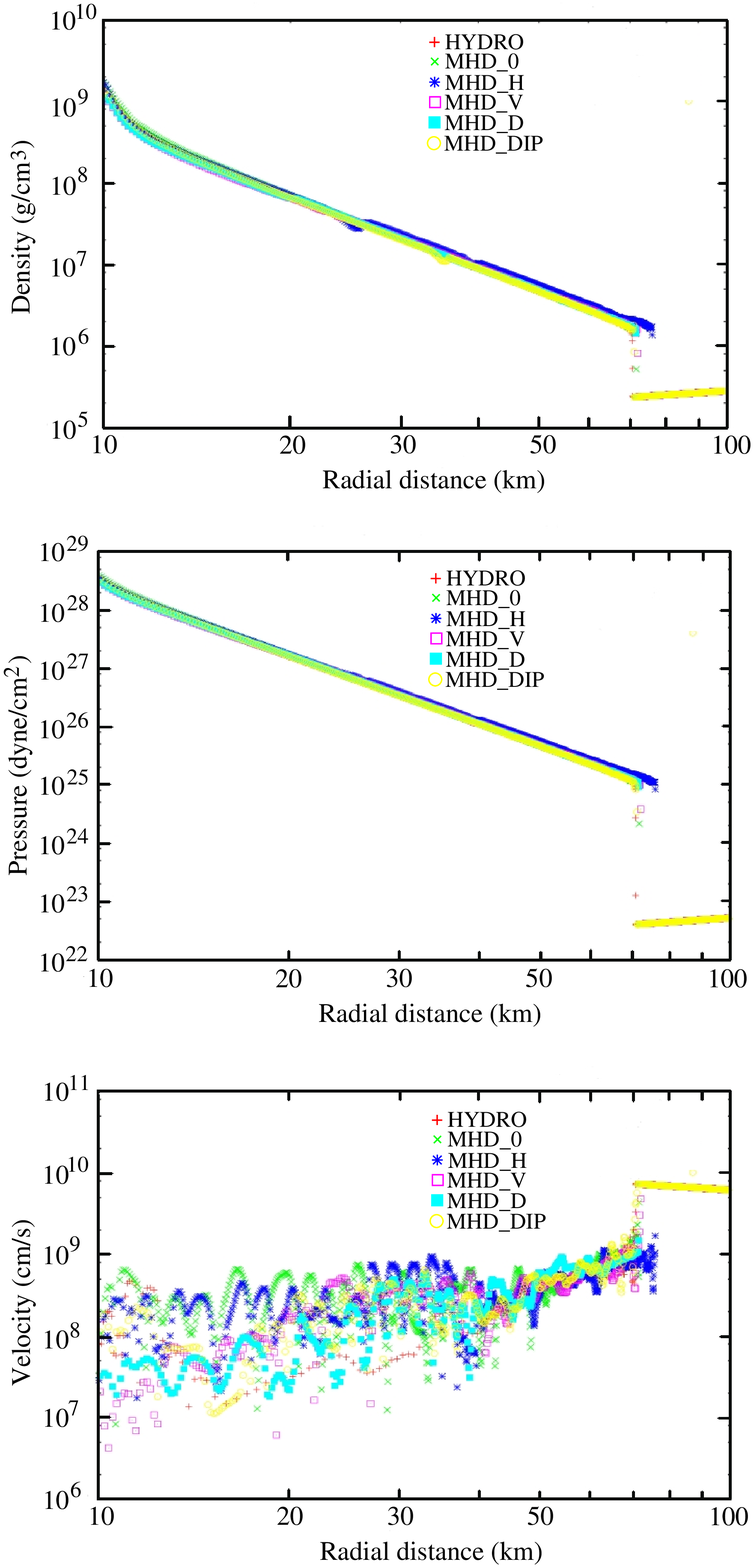}
  \caption{Radial profiles of density (top), pressure (middle) and velocity (bottom) for $\dot{M}=\dot{M}_{0}$ and various field configurations (labeled).}
  \label{Fig:profs2a}
\end{figure}

We now consider the case with non--zero magnetic field strength. Fig. \ref{Fig:profs2a} shows the radial profiles of density, pressure and magnitude of the velocity for $\dot{M}=\dot{M}_{0}$ with several field configurations: null (MHD\_0), constant horizontal (MHD\_H), constant vertical (MHD\_V), constant diagonal (MHD\_D) and dipolar (MHD\_DIP), for comparative effects. The initial intensity of the magnetic field in all cases is $10^{12}\,\mathrm{G,}$ except in the
dipole configuration, where it is $10^{12}$~G at the neutron star surface. We also overplot the hydrodynamical solution for comparison.  Note that the profiles are practically the same at this accretion rate indicating that the magnetic field is not playing an important role as far as the dynamics are concerned. 

\begin{figure}[!t]\centering
  \includegraphics[width=0.49\columnwidth]{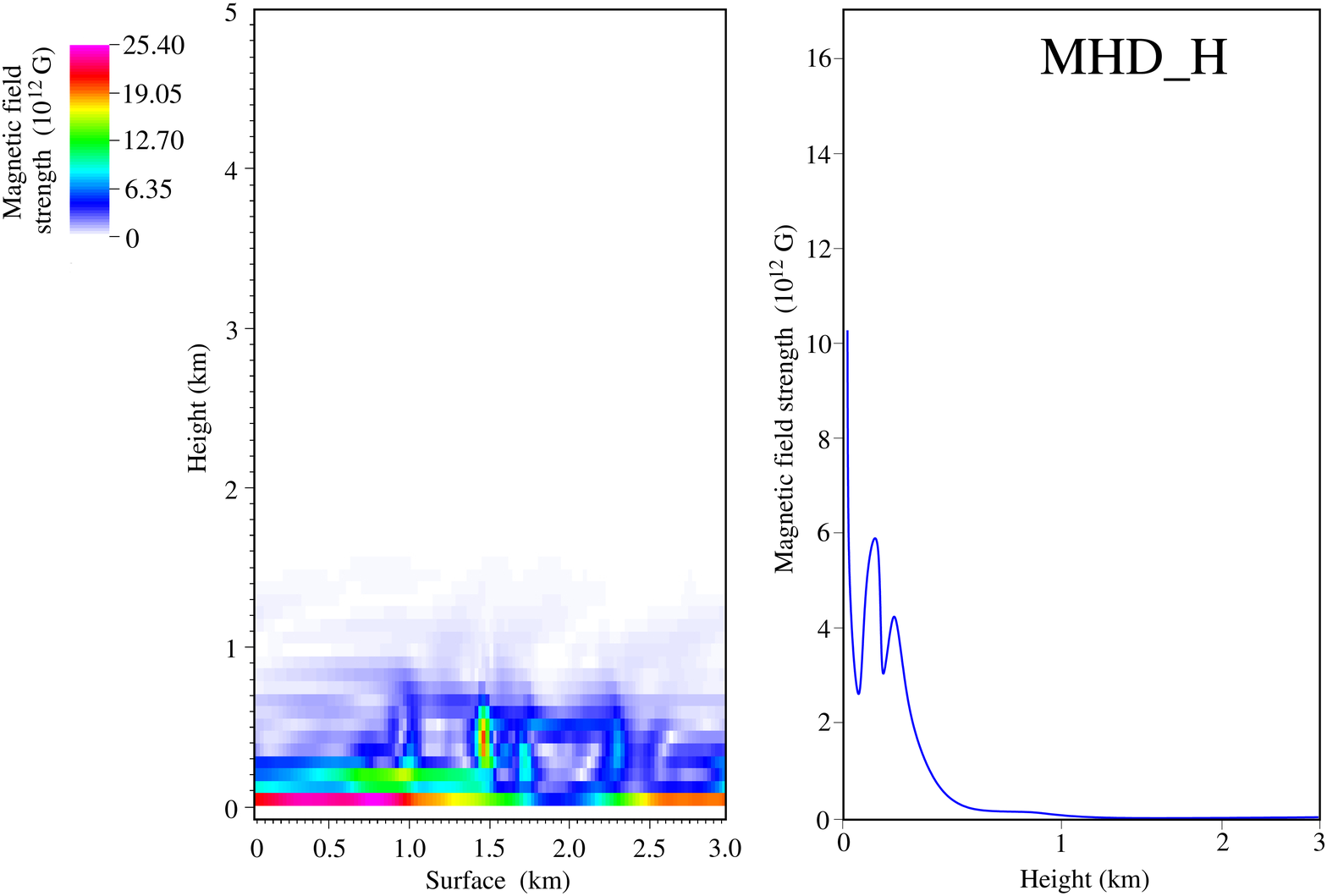}
  \includegraphics[width=0.49\columnwidth]{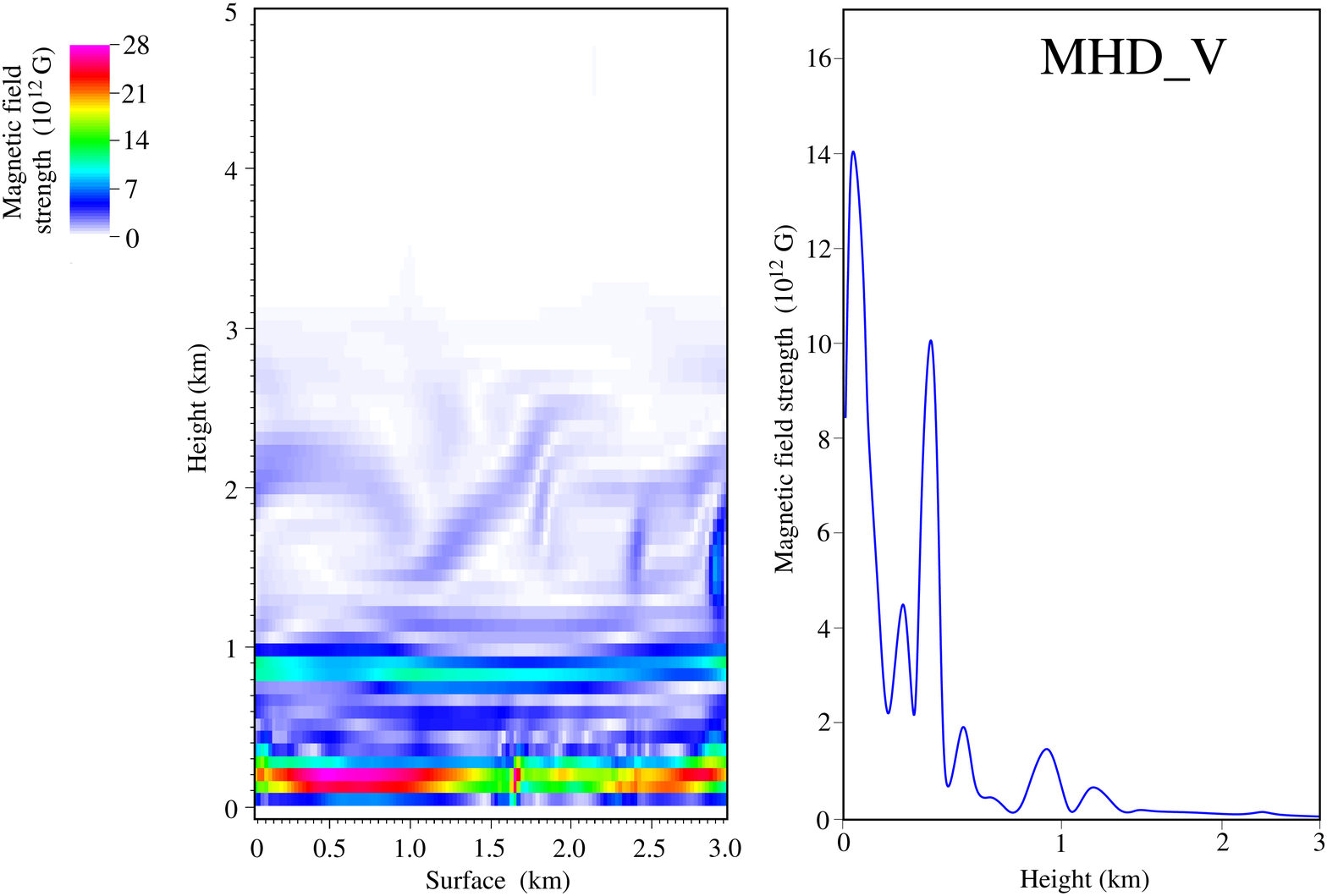}
  \includegraphics[width=0.49\columnwidth]{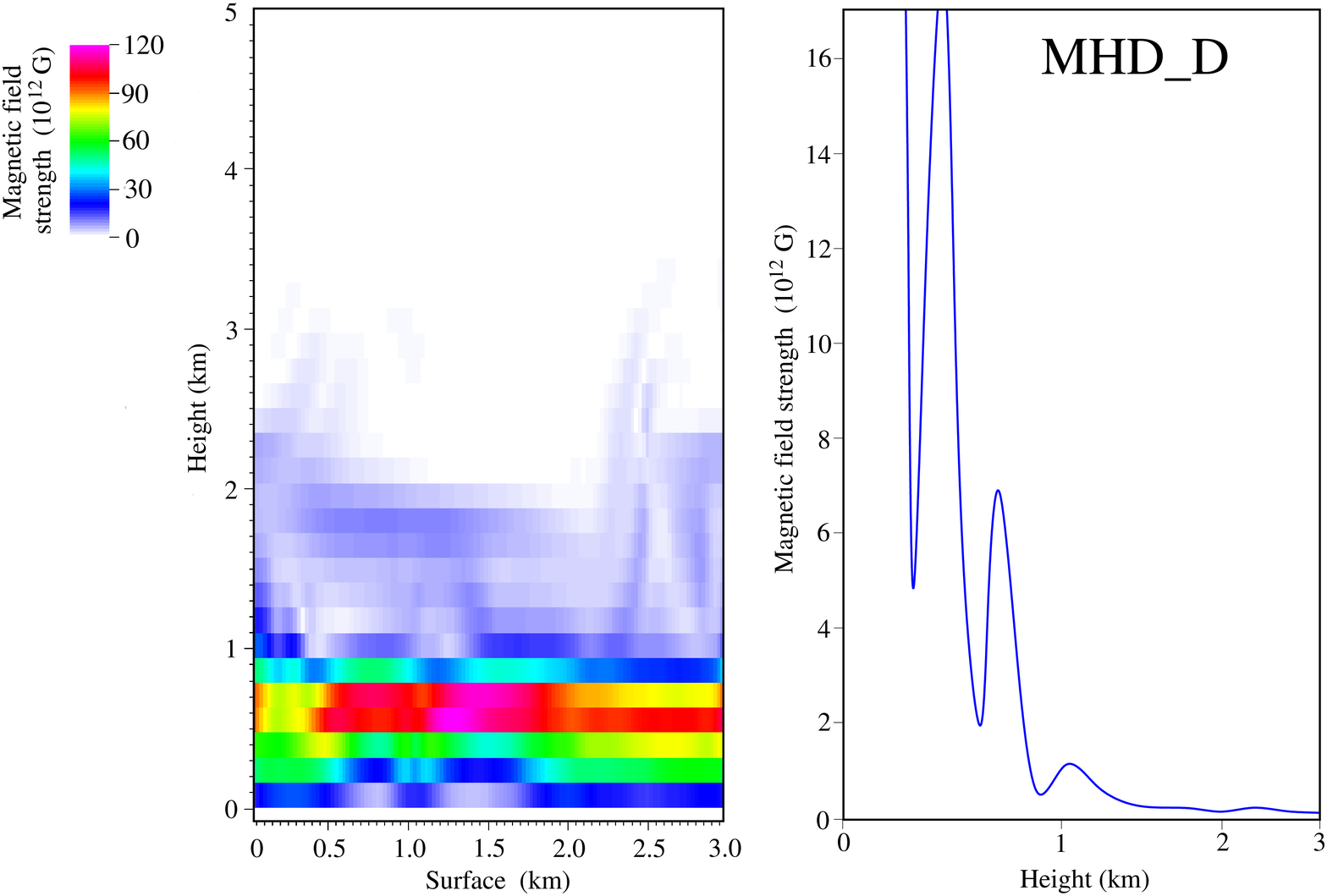}
  \includegraphics[width=0.49\columnwidth]{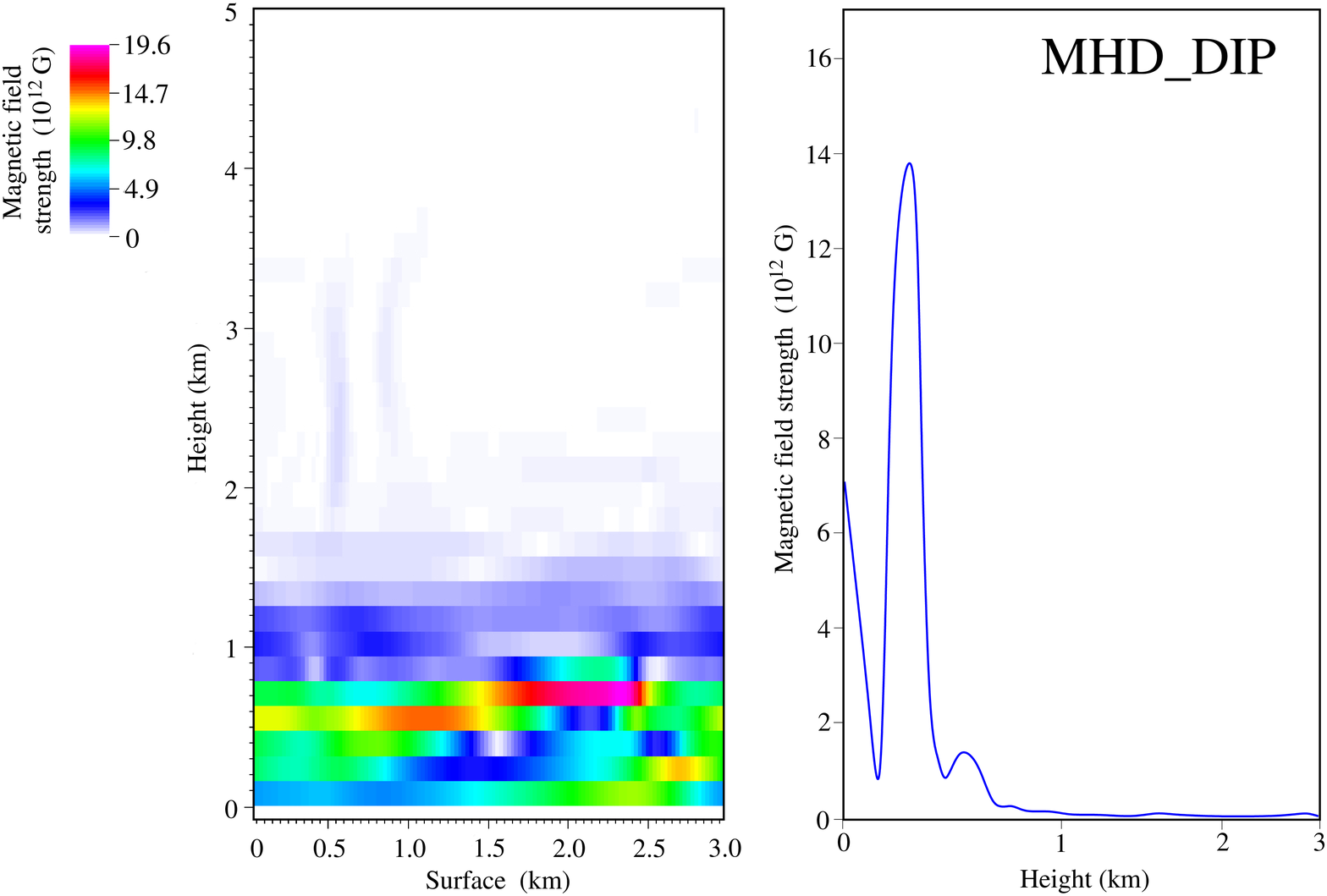}
  \caption{Maps (left half-panel) and horizontally--averaged vertical profiles (right half-panel) of magnetic field strength for a magnetic field that is initially horizontal, case MHD\_H (top left), vertical, case MHD\_V (top right), diagonal, case MHD\_D (bottom left), and dipolar, case MHD\_D (bottom right),     
   after the system has relaxed for $\dot{M}=1000\dot{M}_{0}$. Note the submergence of the magnetic field close to the neutron star surface.}
  \label{Fig:MHD}
\end{figure}

In all simulated cases, regardless of the magnetic field configuration, when the system has relaxed and reached the quasi--stationary state, the field is completely submerged in the neutron star crust. Its intensity rises accordingly, by up to two orders of magnitude for the highest accretion rates. 
Fig. \ref{Fig:MHD} shows the distribution of magnetic field strength after the system has relaxed, when $\dot{M}=1000 \, \dot{M}_{0}$,
for our four initial magnetic field configurations.
It is only within the first km  in the column, where the matter piles up, that the magnetic field is at or above the initial value in the calculation, and the compression is quite clear. 

The initial dynamics in the MHD case are somewhat more violent than in the pure hydrodynamical case. The infalling gas quickly drags the initial field towards the neutron star surface since the ram pressure, $P_{\rm ram}= \rho v^{2}/2$ is substantially greater than the magnetic pressure $P_{\rm mag}=B^{2}/ 8\pi$, even for the smallest accretion rate, $\dot{M}=\dot{M}_{0}$. The increased magnetic pressure as compression takes place is insufficient to overcome this flow, and large field strengths close to the surface result. The effect on the large scale dynamics is thus of a more transitory nature, and sensitive to the initial conditions, than a permanent feature. As a second point, we note that the magnetic field, advected along with the flow, fluctuates in strength strongly in the shocked region as it piles up against the lower boundary, where neutrino cooling is efficient. The additional piling up of material makes it even harder for the field to rise to significant levels as the evolution proceeds further. Nevertheless, as the system
evolves the turbulent structures that form initially begin to smooth
themselves until they disappear completely in the hydrodynamical case, but some small scale structure remains when magnetic fields are present. 

Once the accretion rate drops significantly, it is in principle possible that the field will rise buoyantly through the envelope, playing some dynamical role as the accretion time becomes long and the balance between ram and magnetic pressure is reversed. This will occur on a much longer time scale than simulated here, and its modeling requires a different set of assumptions in terms of the present set of calculations.

\section{Conclusions}
\label{sec:conclusions}

We have presented the results of two--dimensional simulations of accretion in the hypercritical, neutrino--cooled regime onto the surface of a neutron star, using the FLASH code. The flow in accretion columns for a variety of initial accretion rates was simulated until a steady state was reached. We find that at this stage, the location of the accretion shock, where the flow transitions from free fall to subsonic settling onto the neutron star surface, is well reproduced when compared with the analytical estimates of \citet{Chevalier:1989jk}. However, close to the surface, matter piles up, the solution is no longer adiabatic, and the self--similar character of the flow breaks down as expected. 

We performed a detailed comparison of the hydrodynamical and ideal MHD routines in FLASH, and found excellent agreement between the two when the initial field is null. For various finite field configurations (initially horizontal, vertical, diagonal and dipolar), we find that performing the calculations in two dimensions does not allow for any additional buoyancy effects of the field to be manifested: for all accretion rates simulated, the initial field is entirely advected by the flow and submerged close to the neutron star surface. Its intensity rises accordingly, by up to two orders of magnitude in some cases. In principle, thus, it is possible for such an accretion episode following core collapse and the formation of a proto--neutron stars to effectively bury the initial field and delay the appearance of a classical radio pulsar \citep{Muslimov:1995sf}. The simulated time scales at present do not allow us to place hard constraints on the re--diffusion of the field at late times, and a more quantitative estimation of this is left for future work. 

\acknowledgments
CGB acknowledges support from a DGEP--UNAM scholarship. Financial support for this work was provided in part by CONACyT (45845E)
and DGAPA--UNAM (IN 122609). 
The software used in this work was in part developed by the DOE--supported ASC / Alliance Center for Astrophysical Thermonuclear Flashes at the University of Chicago. The numerical calculations were carried out on the KanBalam Supercomputer at DGSCA, UNAM, whose support team is gratefully acknowledged. We thank the anonymous referee for comments and criticism which helped improve this final version.

\bibliography{BLP_RevMex_2}

\end{document}